%% file: main.tex
\documentclass[9pt,twocolumn,twoside]{pnas-new}
\templatetype{pnasresearcharticle} % Two-column research-article template

\articletype{Applied Mathematics}%%%% Article topic/classification

\usepackage{bm}
\usepackage{amsthm}
\usepackage{comment}
\allowdisplaybreaks

\input{arxiv_preamble}
\begin{document}

\title{Inverse origami inspired by insect morphogenesis: geometric design of wrinkle structures for curved surface formation}

% Use letters for affiliations, numbers to show equal authorship (if applicable) and to indicate the corresponding author
\author[a,1]{Kentaro Morikawa}
\author[b]{Yuuki Shimizu}
\author[c]{Nen Saito}
\author[a]{Yasuhiro Inoue}

\affil[a]{Department of Micro Engineering, Graduate School of Engineering, Kyoto University, Kyoto 615-8540, Japan}
\affil[b]{Faculty of Science, Academic Assembly, University of Toyama, Toyama 930-8555, Japan}
\affil[c]{Life Science Center for Survival Dynamics, Tsukuba Advanced Research Alliance (TARA), University of Tsukuba, Tsukuba 305-8577, Japan}

% Please give the surname of the lead author for the running footer
\leadauthor{Morikawa}

% Please add a significance statement to explain the relevance of your work
\significancestatement{Insects face a spatial paradox: a larger post-molt body must be prepared inside a smaller pre-molt body. They solve it by prefabricating a wrinkled surface that unfolds under body-fluid pressure, as in the horn of the rhinoceros beetle. Origami makes shapes by folding flat sheets. Insects do the reverse, unfolding preformed wrinkles, and we call this inverse origami. Inspired by this strategy, we ask: given a target surface, what wrinkle structure unfolds into it? Regarding a wrinkle structure as a stretch field, we derive such a structure from a prescribed surface. This inverse-origami approach provides a geometric framework for designing and fabricating deployable curved structures.}

% Please include corresponding author, author contribution and author declaration information
\authorcontributions{K.M., N.S., and Y.I. designed research; K.M. and Y.S. performed research; K.M. contributed new analytic tools; K.M. analyzed data; and K.M., Y.S., N.S., and Y.I. wrote the paper.}
\authordeclaration{The authors declare no competing interest.}
\correspondingauthor{\textsuperscript{1}To whom correspondence should be addressed. E-mail: morikawa.kentaro.68a@kyoto-u.jp}

% At least three keywords are required at submission. Please provide three to five keywords, separated by the pipe symbol.

\begin{abstract}
    Insect exoskeletons abruptly acquire complex three-dimensional forms during metamorphosis from larva to pupa.
    This transformation occurs through the unfolding of preformed wrinkle structures inside the larval body under internal pressure.
    Inspired by this process, we investigate an inverse-origami strategy for fabricating curved surfaces.
    Whereas conventional origami produces three-dimensional objects by folding a flat sheet, this strategy produces curved surfaces by unfolding preformed wrinkle structures.
    The central design problem is to derive a wrinkle structure from a prescribed target surface.
    Our key idea is to regard a wrinkle structure as a unidirectional stretching field, which specifies at each point the stretch produced when the structure unfolds.
    We first derive such a field from the target surface geometry and then translate it into spatially varying wrinkle directions and spacing, so that the wrinkle structure unfolds into the target.
    The resulting numerical designs approximate the intrinsic metric of the target surface.
    We demonstrate the method on multiple target surfaces through numerical construction, mechanical unfolding simulations, and 3D-printed prototypes.
    This work provides a biologically inspired geometric design principle for encoding prescribed surface shapes in wrinkle structures, as well as a framework for designing and fabricating deployable curved surfaces.
\end{abstract}

\dates{This manuscript was compiled on \today}
\doi{\url{www.pnas.org/cgi/doi/10.1073/pnas.XXXXXXXXXX}}

\maketitle
\thispagestyle{firststyle}
\ifthenelse{\boolean{shortarticle}}{\ifthenelse{\boolean{singlecolumn}}{\abscontentformatted}{\abscontent}}{}

\begin{figure}[htbp]
    \centering
    \includegraphics[width=\linewidth]{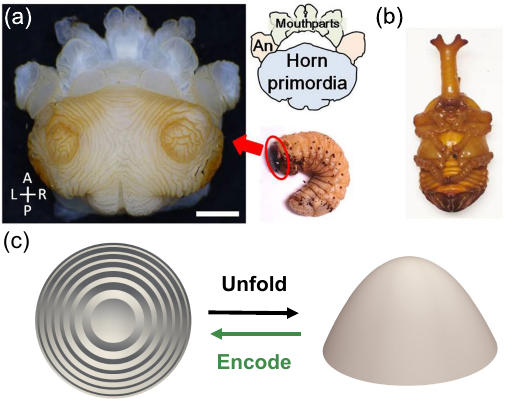}
    \caption{Morphogenesis of the beetle horn and the concept of inverse origami. (a) Wrinkle structure of the horn primordium. The horn is formed as a wrinkle structure inside the larval body, and a three-dimensional curved surface is generated by unfolding it. (b) Shape of the pupa. Panels (a) and (b) are reprinted from~\cite{Matsuda2017} (CC BY 4.0). (c) Concept of inverse origami. A wrinkle structure (left) unfolds into a curved surface (right). The design problem addressed in this study is the reverse operation: encoding a target surface into a wrinkle structure that unfolds into it.}
    \label{fig:beetle_horn}
\end{figure}

% Note (PNAS style): the introduction begins here without an "Introduction" heading.
Insects and other exoskeletal animals grow through repeated molting.
This is a spatially paradoxical process, in which the larger post-molt body must be prepared in advance inside the smaller pre-molt body.
Under the constraint that the shape of the large body must be formed within the small body, insects construct the prospective body-surface shape as a primordium folded into wrinkles,
and complete the final surface shape by unfolding it under body-fluid pressure at molting~\cite{Chapman2012}.
Such morphogenesis mediated by wrinkle structures is a universal strategy widely observed in insect exoskeletal morphogenesis, not only in the beetle horn but also in the unfolding of the \textit{Drosophila} wing~\cite{Hadjaje2024} and the treehopper helmet~\cite{Adachi2020}.
A particularly striking case is when a complex three-dimensional shape, absent in the larval stage, emerges abruptly upon molting. The horn of the Japanese rhinoceros beetle (\textit{Trypoxylus dichotomus}) is a typical example (Fig.~\ref{fig:beetle_horn}).
If the goal were merely to enlarge the body, the existing body surface could serve as a mold for the folded primordium. In the case of the horn, however, the wrinkle structure of the horn primordium is formed inside the larval body, which possesses no horn shape to serve as a mold. The information specifying the horn shape therefore cannot be copied from a mold and must reside in the wrinkle structure itself.

For the beetle horn, the three-dimensional shape after unfolding is already encoded in the wrinkle structure of the primordium.
Two observations support this view: cell proliferation contributes little to the morphological change during unfolding at molting, and a horn primordium excised from the larva and inflated artificially---even after chemical fixation---unfolds into the horn shape~\cite{Matsuda2017}.
How the wrinkle structure itself is formed has been studied extensively, including the active introduction of folds along lines of contractile signal~\cite{Adachi2018}, the genetic control of the fold pattern~\cite{Adachi2020a}, mechanical simulations of fold formation in the primordium~\cite{Inoue2020,Morikawa2022}, and, more broadly, the mechanics of thin-sheet wrinkling and epithelial folding~\cite{Cerda2003,Tobasco2022,Shyer2013}.
These studies address how wrinkles are generated through cellular and mechanical processes.
In contrast, the geometric principle by which a target three-dimensional shape is encoded into a wrinkle structure---that is, which wrinkle structure unfolds into a given surface---remains unresolved.
Taking this biological strategy as inspiration, the present study addresses the corresponding geometric design problem: by what rule can a prescribed surface shape be encoded into a wrinkle structure that unfolds into it?

In engineering, the fabrication of a prescribed curved surface from a planar sheet has been approached in two broad ways, which differ in what is designed from the target shape and in how the shape is realized.
The first is origami design: from the target surface, a crease pattern on a flat, inextensible sheet is designed, and the shape is realized by folding.
This includes universal algorithms that construct a crease pattern folding into an arbitrary target shape~\cite{Demaine2017}, methods that perturb tessellations within a differential-geometric framework to derive creases fitting a target surface~\cite{Sardas2025}, and methods that approximate a target surface by optimizing or constructing tessellation patterns~\cite{Dudte2016,Zhao2023}.
The second is programmable materials, which shape a planar sheet through an encoded field: from the target surface, a field on the flat sheet is designed---the growth field of anisotropically growing bilayers~\cite{Rees2017}, the director field of liquid-crystal elastomers~\cite{Aharoni2018}, the fusing pattern of sheets that unfold by in-plane inflation~\cite{Panetta2021}, or the areal-growth-ratio distribution on heat-shrink films~\cite{Morikawa2026}---and the shape is realized by the sheet deforming according to that field.
In both approaches the starting point is a flat sheet: the target surface is encoded as a crease pattern or a field on the flat sheet, which is then folded or deformed into the target.

The insect strategy differs from both approaches in its starting point.
The starting point is not a flat sheet but an already folded wrinkle structure, and the target surface is realized by unfolding this structure.
Insects thus create curved surfaces by unfolding preexisting folds (Fig.~\ref{fig:beetle_horn}(c)). This is the reverse of origami, and we call it inverse origami.
The corresponding design problem is to derive the wrinkle structure, including its crease pattern, that unfolds into a given target surface (``Encode'' in Fig.~\ref{fig:beetle_horn}(c)).
The key viewpoint of the present study is that this wrinkle structure can itself be regarded as a field on the plane: unfolding stretches the wrinkles perpendicular to their ridges by a ratio set by their fineness, so that the wrinkle structure represents the local stretch at each point.
Through this viewpoint, the design problem takes the same form as for programmable materials---encoding a target surface into a field on the plane---even though the starting configuration is not flat.

Based on this viewpoint, we formulate the wrinkle structure as a vector field in which deformation near each point is a unidirectional stretch (a unidirectional stretching field), and provide a geometric principle for solving the inverse problem of deriving a wrinkle structure that unfolds isometrically into a prescribed target surface. We then present a numerical method for constructing wrinkle structures from general target surfaces, quantify their deviations from exact isometry, and further examine their unfolding behavior through simulations. We also demonstrate the physical feasibility of the approach through proof-of-concept fabrication. Together, these results establish inverse origami as a geometric design principle for wrinkle structures that unfold into prescribed curved surfaces, and offer a geometric hypothesis for how insects may encode three-dimensional shapes in wrinkle structures.

\section*{Inverse origami principle}
\subsection*{The relationship between wrinkle structures and surfaces}
\begin{figure}[htbp]
    \centering
    \includegraphics[width=\linewidth]{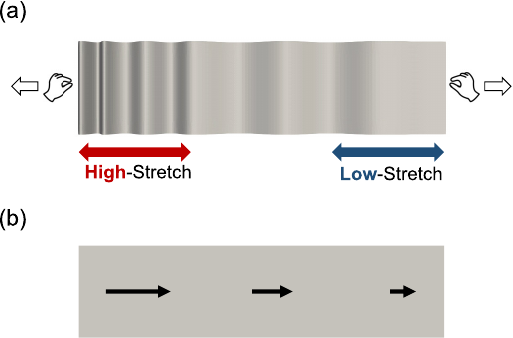}
    \caption{Regarding a wrinkle structure as a vector field of stretch. (a) A wrinkle structure and a surface shape are connected by the operation of stretching the wrinkles. The direction in which the wrinkles stretch is restricted to the direction perpendicular to the wrinkle ridges, and the magnitude of the stretch corresponds to the fineness of the wrinkles. (b) The vector field of wrinkle stretch corresponding to the wrinkle structure in (a). The length of each arrow represents the magnitude of the stretch, and its direction represents the direction of the stretch.}
    \label{fig:concept}
\end{figure}

To formulate the encoding principle geometrically by which a target surface shape is encoded into a wrinkle structure, we first consider the relationship between the wrinkle structure and the surface.
A wrinkle structure and a surface shape are connected by the operation of stretching the wrinkles (Fig.~\ref{fig:concept}(a)).
This stretching is not a stretching of the surface itself, but a stretch in the sense of isometrically unfolding the folds.
The direction in which the wrinkles stretch is restricted to the direction perpendicular to the wrinkle ridges.
Moreover, under the assumption that the wrinkle amplitude is constant, the magnitude of the stretch corresponds to the fineness of the wrinkles.
That is, a wrinkle structure can be regarded as a vector field of deformation in which the deformation in the neighborhood of each point is a unidirectional stretch (Fig.~\ref{fig:concept}(b)).
In other words, this vector field captures, as a literal stretch, the stretch that arises from the isometric unfolding of the wrinkles.
We refer to this vector field as a unidirectional stretching field.
In the present study, for simplicity, we treat only wrinkle structures on the plane.
Accordingly, the unidirectional stretching field is a vector field $\bm{s}(u,v)$ on the plane.

To examine the relationship between the wrinkle structure and the unidirectional stretching field, we first describe the shape of the wrinkle structure using the terminology of plane waves. That is, using a periodic function $f(\phi)$ with period $2\pi$ and amplitude $1$, we describe the wrinkle structure as a height function $w(u,v)$ on the $(u,v)$ plane,
\begin{equation}
    \label{eq:w_phi}
    w(u,v) = A f(\phi(u,v))
\end{equation}
where $A$ is the amplitude of the wrinkles and $\phi(u,v)$ is the phase of the wrinkles.
In general, the wrinkle amplitude $A$ may depend on position; here, however, we take it to be constant and formulate the magnitude of the stretch as depending only on the fineness of the wrinkles.
The direction in which the phase of a wave advances and its rate are, in general, described by the wave-number vector $\bm{k}$:
\begin{equation}
    \label{eq:k_phi}
    \nabla \phi(u,v) = \bm{k}(u,v)
\end{equation}
The direction $\hat{\bm{s}}(u,v)=\bm{s}(u,v)/\|\bm{s}(u,v)\|$ of the stretch vector at each point $(u,v)$ of the unidirectional stretching field is perpendicular to the wrinkle ridges, which, in wave terminology, lies along the direction of the wave-number vector. Furthermore, the fineness of the wrinkles (which is related to the magnitude of the stretch $\|\bm{s}(u,v)\|$) corresponds to the magnitude of the wave-number vector.

\begin{figure}[htbp]
    \centering
    \includegraphics[width=0.6\linewidth]{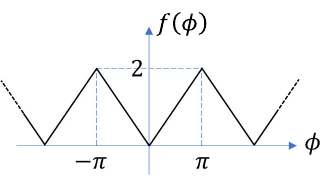}
    \caption{The piecewise-linear periodic function $f(\phi)$. Using this function, one can represent wrinkles with sharp creases rather than smooth bends, which makes it possible to design isometric wrinkle structures.}
    \label{fig:piecewise_linear}
\end{figure}

To consider the relationship between the magnitude of the stretch vector and the magnitude of the wrinkle wave-number vector, we need to specify the periodic function $f(\phi)$.
In the present study, we take $f(\phi)$ to be the piecewise-linear function given, in the range $-\pi < \phi \leq \pi$, by
\begin{equation}
    \label{eq:f_phi}
    f(\phi) = \frac{2}{\pi}|\phi| \quad \text{if } -\pi < \phi \leq \pi
\end{equation}
(Fig.~\ref{fig:piecewise_linear}).
We use this function in order to be able to represent sharp creases.
Our goal is to design a wrinkle structure that is isometric to the target surface, and if only smooth folds are allowed, isometric folding is difficult. This can be understood by considering the problem of folding a hemisphere.
A hemisphere is a surface whose Gaussian curvature is positive and constant everywhere; if, for example, it is deformed by pushing the pole inward, a circular fold arises, and if this fold is a smooth bend, a region of negative Gaussian curvature appears. By Gauss's \textit{Theorema Egregium}, which states that the Gaussian curvature is preserved at every point under an isometric deformation, we see that isometry is broken in this case.
If, on the other hand, the folding is done with sharp creases, one can fold while preserving the Gaussian curvature everywhere except on the crease curves.
On the basis of this consideration, we decided to use a piecewise-linear function as $f$ in order to handle sharp creases.
Note that, even though $f$ is piecewise linear, a smoothly varying phase field $\phi(u,v)$ yields a wrinkle surface that is smooth between the crease lines and nonsmooth only along the level sets where $f$ changes slope.
Thus, the piecewise-linear choice of $f$ allows spatially varying, curved wrinkle patterns while retaining sharp creases.

Using the function in Eq.~(\ref{eq:f_phi}), it follows from Eq.~(\ref{eq:k_phi}) that the slope of the wrinkle structure $\|\nabla w(u,v)\|=A\|\nabla f(\phi(u,v))\|$ is
\begin{equation}
    \label{eq:grad_w}
    \|\nabla w(u,v)\| = \frac{2A}{\pi} \|\bm{k}(u,v)\|
\end{equation}
Because stretching the wrinkles is an operation that changes their slope and locally flattens them, by the Pythagorean theorem the stretch ratio $\|\bm{s}(u,v)\|$ is
\begin{equation}
    \label{eq:s_w}
    \|\bm{s}(u,v)\| = \sqrt{1 + \|\nabla w(u,v)\|^2}
\end{equation}
From Eqs.~(\ref{eq:grad_w}) and~(\ref{eq:s_w}), we have $\|\bm{k}\|=\frac{\pi}{2A}\sqrt{\|\bm{s}\|^2-1}$, and since $\bm{s}$ and $\bm{k}$ have the same direction, as noted above, the relationship between the unidirectional stretching field and the wrinkle wave-number vector is
\begin{equation}
    \label{eq:k_s}
    \bm{k}(u,v) = \frac{\pi}{2A}\sqrt{\|\bm{s}(u,v)\|^2-1} \hat{\bm{s}}(u,v)
\end{equation}
By Eqs.~(\ref{eq:w_phi}), (\ref{eq:k_phi}), (\ref{eq:f_phi}), and~(\ref{eq:k_s}), the wrinkle structure can be described as a unidirectional stretching field.

\subsection*{Encoding target surfaces into wrinkle structures}
\begin{figure*}[t]
    \centering
    \includegraphics[width=\linewidth]{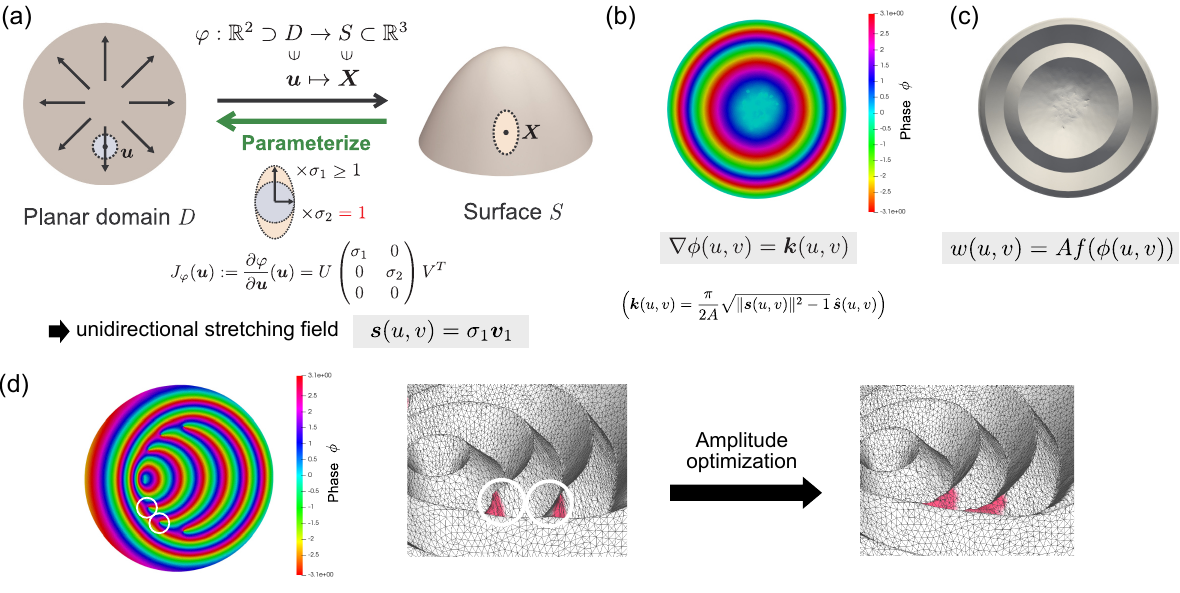}
    \caption{Overview of the encoding of a target surface into a wrinkle structure. (a) Conversion of the target surface into a unidirectional stretching field (schematic). For the deformation to be represented by a vector field, the deformation in the neighborhood of each point must be unidirectional; that is, one of the singular values of the Jacobian matrix $J_\varphi$ must be equal to $1$. For the unidirectional deformation to be a stretch rather than a contraction, the other singular value must be greater than or equal to $1$. The stretching field $\bm{s}$ is then given by Eq.~(\ref{eq:s_param}). (b) The phase field $\phi$ obtained from the stretching field through the wave-number field $\bm{k}$ (Eqs.~(\ref{eq:k_s}) and~(\ref{eq:k_phi})), computed for the paraboloid target in (a). (c) The constant-amplitude wrinkle structure $w=Af(\phi)$ (Eqs.~(\ref{eq:w_phi}) and~(\ref{eq:f_phi})) obtained from the phase field in (b). (d) Amplitude optimization to resolve the discontinuity of the height function at the singular points of the phase field. Left: the phase field computed for the overhang shape (third from the left in Fig.~\ref{fig:demonstration}(a)) with $A=0.08$. Singular points at which the phase is discontinuous appear; two of them are circled. Middle: the constant-amplitude wrinkle structure around the two circled singular points, with the regions around the singular points colored. The height function is discontinuous there, and the mesh is stretched by the discontinuity. Right: the same part of the wrinkle structure after the amplitude optimization, with the same regions colored; the discontinuity is resolved.}
    \label{fig:parameterize}
\end{figure*}
Focusing on the fact that a wrinkle structure and a surface shape are connected by the operation of stretching the wrinkles, we formulated above the relationship between the stretching operation and the wrinkles.
Therefore, to encode a target surface shape into a wrinkle structure, it suffices to convert the target surface shape into a unidirectional stretching field.
Mapping a surface $S \subset \mathbb{R}^3$ onto a planar domain $D \subset \mathbb{R}^2$ is an operation called parameterization.
The stretch in the neighborhood of each point in the deformation $\varphi: D \to S$ from $D$ to $S$ is represented by the Jacobian matrix $J_\varphi(u,v):=\frac{\partial \varphi}{\partial \bm{u}}$ of $\varphi$.
In the singular value decomposition of $J_\varphi(u,v)$,
\begin{equation}
    \label{eq:svd}
    J_\varphi(u,v) = U \begin{pmatrix} \sigma_1 & 0 \\ 0 & \sigma_2 \\ 0 & 0 \end{pmatrix} V^T
\end{equation}
the singular values $\sigma_1, \sigma_2$ represent the stretch ratios along the principal directions $\bm{v}_1, \bm{v}_2$, respectively. The principal directions are the column vectors of $V=\begin{pmatrix} \bm{v}_1 & \bm{v}_2 \end{pmatrix}$.
That the deformation is a unidirectional stretch in the neighborhood of each point means that $\sigma_1 \geq 1$ and $\sigma_2 = 1$.
Therefore, to obtain a unidirectional stretching field, it suffices to parameterize the target shape so that the singular values of the Jacobian matrix satisfy $\sigma_1 \geq 1$ and $\sigma_2 = 1$.
In this parameterization, the direction of the stretch is the $\bm{v}_1$ direction and the stretch ratio is $\sigma_1$.
Therefore, the unidirectional stretching field is given by
\begin{equation}
    \label{eq:s_param}
    \bm{s}(u,v) = \sigma_1 \bm{v}_1
\end{equation}
The concrete construction of this parameterization is described in the next section.

With the above, we have geometrically formulated the principle by which a target surface shape is encoded into a wrinkle structure on the plane.
This encoding procedure can be summarized as follows.
First, for the target surface shape $S$, we parameterize it so that the singular values of the Jacobian matrix of $\varphi: D \to S$ satisfy $\sigma_1 \geq 1$ and $\sigma_2 = 1$ (Fig.~\ref{fig:parameterize}(a)).
From that parameterization, we obtain the unidirectional stretching field $\bm{s}(u,v)$ by Eq.~(\ref{eq:s_param}), and the wrinkle wave-number vector field $\bm{k}(u,v)$ by Eq.~(\ref{eq:k_s}).
From $\bm{k}(u,v)$, we determine $\phi(u,v)$ satisfying Eq.~(\ref{eq:k_phi}), that is, the scalar potential (Fig.~\ref{fig:parameterize}(b)).
The method for carrying out this integration is described in the next section.
Finally, we obtain the wrinkle height function $w(u,v)$ by Eq.~(\ref{eq:w_phi}) (Fig.~\ref{fig:parameterize}(c)).

\subsection*{Verification of isometry via the induced metric}
We verify that the wrinkle structure obtained by the encoding method based on the inverse origami principle described above is isometric to the target surface shape.
An isometric deformation is a deformation that preserves the metric tensor.
We therefore compare the metric that the target surface induces on the planar domain $D$ through the parameterization $\varphi$ with the metric induced by the wrinkle surface $\varphi_\mathrm{W}(u,v)=(u,v,w(u,v))$.
Computing the two pullback metrics and equating them yields the condition (see \textit{SI Appendix} for the derivation)
\begin{equation}
    \label{eq:isometry_condition}
    \nabla\phi = \pm\frac{\pi}{2A}\sqrt{(\sigma_1)^2 - 1}\;\bm{v}_1
\end{equation}
From Eq.~(\ref{eq:s_param}), since $\sigma_1=\|\bm{s}\|$ and $\bm{v}_1=\hat{\bm{s}}$, this condition coincides with Eqs.~(\ref{eq:k_phi}) and~(\ref{eq:k_s}) used in the encoding, up to the double sign.
The double sign corresponds to reversing the direction of the stretch; however, since the notion of stretch is symmetric under a $180^\circ$ rotation, the two choices are physically equivalent.
Therefore, suppose that the target surface admits a parameterization satisfying $\sigma_1\geq1$ and $\sigma_2=1$ exactly, and that the associated wave-number field admits a globally compatible phase field satisfying $\nabla\phi=\bm{k}$ after a consistent choice of orientation, with the necessary compatibility conditions satisfied around singularities and non-simply-connected regions. For constant amplitude $A$, the encoding method then yields a wrinkle surface with the same induced metric as the target surface wherever $f$ is differentiable. Along the crease curves, where $f$ is not differentiable, the isometry is understood in the piecewise-smooth sense.

\section*{Numerical construction of wrinkle structures for general surfaces}
We propose a numerical method for encoding a target surface shape into a wrinkle structure.
The encoding is carried out by converting the target surface shape into a unidirectional stretching field and then constructing the wrinkle structure from that unidirectional stretching field.
\subsection*{Parameterization for converting a surface into a unidirectional stretching field}
% Local/Global algorithmを用いる
As a method for numerically obtaining a parameterization that yields the desired Jacobian matrix $J_\varphi$, we use the local/global algorithm~\cite{Liu2008}. The local/global algorithm extends the framework proposed in as-rigid-as-possible surface modeling~\cite{Sorkine2007} and applies it to parameterization; for a triangle mesh $S$, it maps $S$ onto a planar domain $D$ in such a way that the Jacobian matrix $J_i(u,v)$ of each triangle $i$ in the deformation $\varphi: D \to S$ is constrained to a set $M$ of allowed linear transformations. In the present study, the set of allowed linear transformations is as follows.
\begin{equation}
    \label{eq:allowed_linear_transformations}
    M = \left\{ L \in \mathbb{R}^{3 \times 2} \mid L = U\begin{pmatrix} \sigma_{1} & 0 \\ 0 & 1 \\ 0 & 0 \end{pmatrix}V^T, \sigma_{1} \geq 1 \right\}
\end{equation}
where $U, V$ are orthogonal matrices and $\sigma_{1}$ is the larger singular value of $L$; this constraint enforces $\sigma_1\geq1$ and $\sigma_2=1$, i.e., a unidirectional stretch in the neighborhood of each point.
The parameterization is obtained by finding the vertex positions $\{\bm{u}_i\}$ and the per-triangle linear transformations $\{L_i\}$ that minimize the associated as-rigid-as-possible energy, alternating a local phase (which fixes $\{\bm{u}_i\}$ and finds $\{L_i\}$) and a global phase (which fixes $\{L_i\}$ and finds $\{\bm{u}_i\}$), with the initial values given by a conformal parameterization~\cite{Eck1995, Levy2002} (see \textit{SI Appendix} for the energy function and the update rules).
From the resulting parameterization, the unidirectional stretching field $\bm{s}_i=\sigma_{1, i}\bm{v}_{1,i}$ is obtained for each triangle (see Eq.~(\ref{eq:s_param})).

\subsection*{Constructing the wrinkle structure from the unidirectional stretching field}
From the unidirectional stretching field $\bm{s}_i$ obtained above, we first compute the wave-number vector field $\bm{k}_i$ for each triangle by Eq.~(\ref{eq:k_s}).
We then seek $\phi_i$ satisfying Eq.~(\ref{eq:k_phi}).
However, the wave-number vector field $\bm{k}_i$ is not necessarily curl-free, and where its curl is nonzero, no globally continuous scalar function has $\bm{k}_i$ as its gradient.
The stripe pattern method~\cite{Knoppel2015} handles such non-integrable input by representing the phase as the argument $\arg \psi$ of a complex function $\psi$. Ideally, we seek $\psi$ such that
\begin{equation}
    \nabla (\arg \psi) = \bm{k}
\end{equation}
holds. For a non-integrable field $\bm{k}$, however, this relation cannot in general be satisfied exactly. The method instead minimizes its $L^2$ residual while allowing zeros of $\psi$, which introduce singularities where needed to improve alignment with $\bm{k}$. Solving the corresponding discrete minimization---a generalized eigenvalue problem for the nodal values $\{\psi_i\}$, obtained via the inverse power method (see \textit{SI Appendix})---yields the phase $\phi_i=\arg\psi_i$, from which the height $w_i$ at each vertex follows by Eqs.~(\ref{eq:f_phi}) and~(\ref{eq:w_phi}).
Here, the isolines of phase $\phi=0, \pi$ become crease lines; when such a line crosses the interior of a triangle of the mesh, we perform remeshing by adding edges so that the crease lines lie on the edges of the triangle mesh.
Through the above, the mesh of the wrinkle structure can be constructed.

\subsection*{Amplitude optimization to resolve discontinuities at singular points}
Using the numerical construction method described above, one can design a wrinkle structure based on the inverse origami principle.
However, depending on the target surface shape, singular points at which the direction is undetermined may appear in the wave-number vector field, and at those points the phase becomes discontinuous, and consequently the height function becomes discontinuous.
For example, for the overhang shape (third from the left in Fig.~\ref{fig:demonstration}(a)), the numerical construction method above yields a phase such as that in Fig.~\ref{fig:parameterize}(d), and the phase becomes discontinuous at singular points, two of which are circled in the figure. These points are locations where a wrinkle ridge terminates, which appears as a bifurcation of the adjacent ridge, and there the height function becomes discontinuous (Fig.~\ref{fig:parameterize}(d), middle).
In nature as well, locations where wrinkle ridges terminate do exist, as can be seen, for example, in the many such points in Fig.~\ref{fig:beetle_horn}(a).
In natural wrinkles, however, the height function is not discontinuous at the locations where the wrinkle ridges terminate.
Therefore, it should be possible to design a wrinkle structure whose height function is continuous even at the singular points.
In the theoretical treatment of our inverse origami principle, the wrinkle structure was designed under the assumption that the wrinkle amplitude is constant; to resolve the discontinuity at the singular points, we determine the final wrinkle structure by adjusting the wrinkle amplitude as a post-processing step (Fig.~\ref{fig:parameterize}(d)).

The amplitude adjustment uses the same as-rigid-as-possible surface modeling framework as the local/global algorithm~\cite{Sorkine2007}: it seeks the height coordinates $\{w_i\}$ of the vertices while restricting the per-triangle deformation to rigid rotations, so as to remain isometric to the target surface (see \textit{SI Appendix} for the energy function and the update rules).
Through this iteration, the height function $w$ of the wrinkle structure is updated, and a wrinkle structure with position-dependent amplitude $A(u,v)=w(u,v)/f(\phi(u,v))$ is obtained.
As a result, the discontinuity of the height function is alleviated, and a wrinkle structure such as that in Fig.~\ref{fig:parameterize}(d), right, is obtained.
Although this adjustment departs from the constant-amplitude assumption under which the wave-number field was derived, it does not compromise the goal that the assumption served: the adjustment minimizes the same deviation-from-isometry energy directly over the vertex heights, and therefore corrects not only the discontinuities at the phase singularities that motivate it, but also the distributed residual error that the constant-amplitude construction leaves elsewhere, where the phase field satisfies $\nabla\phi=\bm{k}$ only approximately. As quantified in the Demonstrations section, the optimization reduces the deviation from isometry over most of the surface area for every target shape (Fig.~\ref{fig:quantitative}(a) and (b)).
Figure~\ref{fig:parameterize}(d) shows, for the overhang shape, the wrinkle structure designed with amplitude $A=0.08$ and the result of applying the amplitude adjustment to it; these are presented in order to illustrate the necessity of the amplitude optimization. The main results of the wrinkle-structure design method based on the inverse origami principle proposed in this study are presented in the Demonstrations section.

\section*{Evaluation methods for the designed wrinkle structures}
\subsection*{Evaluation of isometry}
The inverse origami principle proposed in this study is intended to yield a wrinkle structure that is isometric to the target surface shape.
As a metric for evaluating how isometric the designed wrinkle structure is to the target surface shape, we use, for each triangle $i$ of the triangle mesh, the Frobenius norm of the difference between the Jacobian matrix $J_i$ and the rigid rotation $R_i$ closest to it.
\begin{equation}
    \label{eq:discrepancy_metric}
    \begin{split}
    D_i &= \|J_i(w) - R_i\|^2_F \\
    &= (\sigma_{1,i} - 1)^2 + (\sigma_{2,i} - 1)^2
    \end{split}
\end{equation}
For summary statistics we also report the equivalent strain-like error
\begin{equation}
    \label{eq:eps_metric}
    \varepsilon_i = \max\left(|\sigma_{1,i} - 1|,\ |\sigma_{2,i} - 1|\right),
\end{equation}
the largest deviation of the principal stretch ratios from unity, which can be read directly as a percentage strain.
Statistics of $\varepsilon_i$ are computed with area weighting over the triangles of the reference surface; because a few nearly degenerate triangles at the phase singularities carry extreme errors while holding negligible area, we report the area-weighted median and 95th percentile, which are robust to these outliers, rather than the mean (see \textit{SI Appendix}).

\subsection*{Unfolding simulation of the wrinkle structure}
Whereas the metric above evaluates the designed structure itself, we also perform an unfolding simulation to test that the wrinkle structure can be mechanically unfolded; the agreement between the unfolded shape and the target is then quantified as described in the next subsection.
Mimicking the pressure-driven unfolding of the beetle horn, the unfolding simulation is a mechanical computation in which internal pressure is applied while isometry is maintained.
The mechanical computation is performed as a simulated-annealing minimization of a discrete shell energy composed of area, interior-angle, dihedral-angle (bending), and internal-pressure terms~\cite{Morikawa2023, Grinspun2003, Tamstorf2013}.
The area and interior-angle terms act so as to maintain isometry, the dihedral-angle (bending) term drives the unfolding of the wrinkles, and the internal-pressure term---which acts only during the first $n_1$ steps and is switched off thereafter---inflates the structure toward a target lumen volume.
The energy functional and the simulation parameters are given in \textit{SI Appendix}.

\subsection*{Evaluation of agreement between the unfolded shape and the target}
To quantify how closely the unfolded structure approximates the target surface, we rigidly register the final state of the unfolding simulation onto the target surface by point-to-point iterative closest point (ICP) alignment, without any rescaling, and measure the unsigned point-to-surface distances between the two surfaces, sampled densely in both directions and normalized by the diagonal of the target's bounding box (see \textit{SI Appendix} for the registration and sampling parameters).

\subsection*{Physical implementation by 3D printing}
In addition to the simulation described above, we physically implemented the wrinkle structures for all five target shapes (Fig.~\ref{fig:demonstration}(a)) by multimaterial 3D printing. Because the designed wrinkle structure is a triangle mesh without thickness, we assigned it a finite, uniform thickness, thinned only in narrow bands along the crease lines so that the creases act as flexible hinges, and printed the regions that must invert during unfolding with a soft material and the remaining regions with a stiff material. The printer, materials, and fabrication parameters are described in \textit{SI Appendix}. The printed structures were unfolded manually.

\section*{Demonstrations}
\begin{figure*}[htbp]
    \centering
    \includegraphics[width=0.8\linewidth]{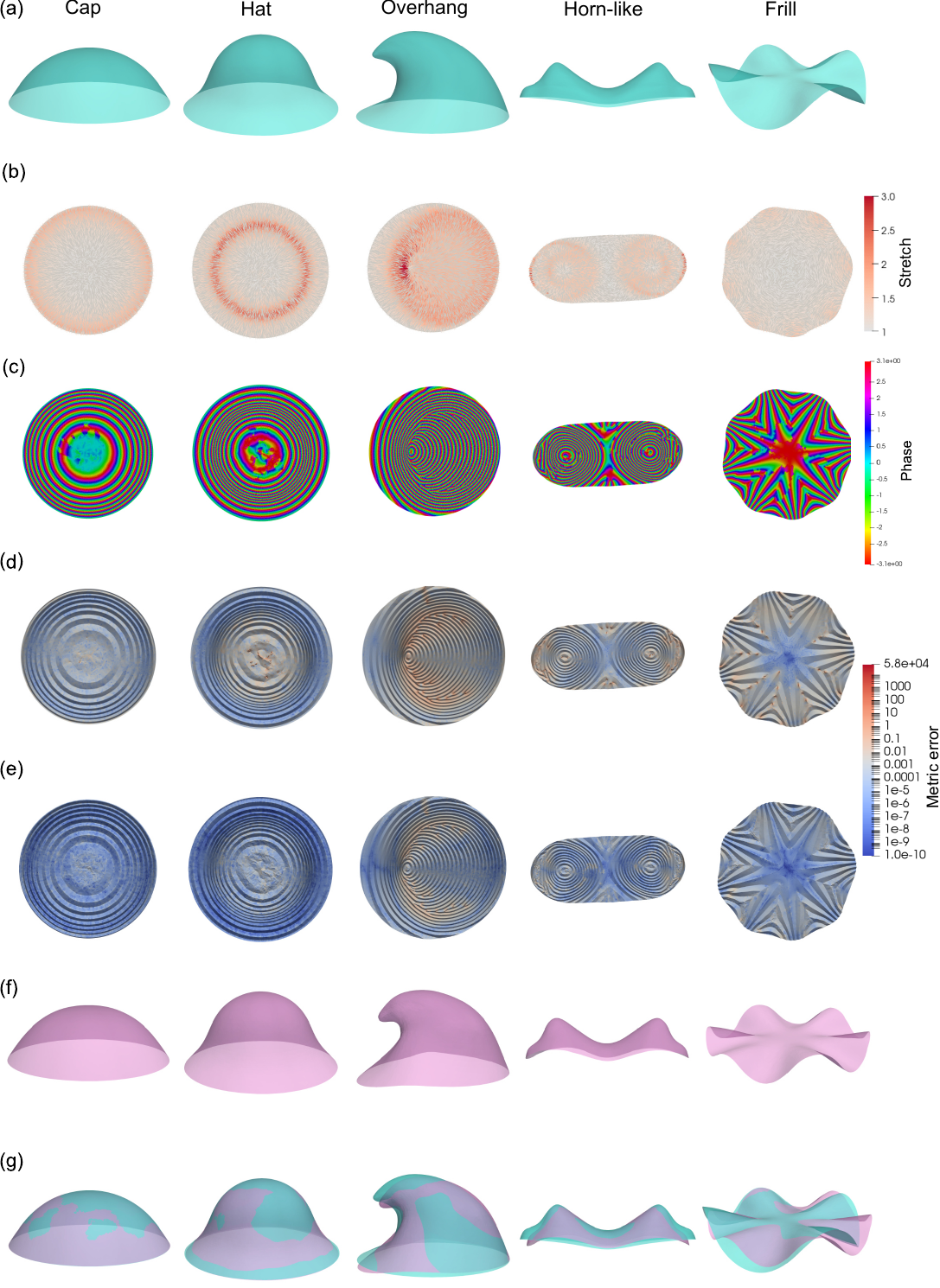}
    \caption{Design results of wrinkle structures obtained by implementing the inverse origami principle, and their evaluation. (a) Target surface shapes. (b) Results of the conversion into the unidirectional stretching field. (c) Results of the computation of the wrinkle phase. (d) Design results of the wrinkle structures. The color represents the deviation from isometry, and the color bar is on a logarithmic scale. (e) Results of the amplitude optimization. The color bar is the same as in (d). (f) Results of the unfolding simulation of the wrinkle structures. (g) The target surface shape and the unfolding-simulation result superimposed. Blue indicates the target surface shape, and pink indicates the unfolding-simulation result.}
    \label{fig:demonstration}
\end{figure*}

We applied the encoding of target surface shapes into wrinkle structures by the inverse origami principle to the cap, hat, overhang, horn-like, and frill target surface shapes shown in Fig.~\ref{fig:demonstration}(a). These targets were chosen to span qualitatively different geometries: dome-like surfaces (cap and hat), a surface with an overhanging region (overhang), a biologically motivated shape (horn-like), and a surface with a wavy, negatively curved rim (frill). The horn-like shape was created with 3D CAD to mimic the shape of the tip of the beetle horn. The wrinkle structure visible in Fig.~\ref{fig:beetle_horn}(a) corresponds to the tip of the horn, whereas the shaft of the horn corresponds to the wrinkle structure that continues on the back side, hidden from the front~\cite{Matsuda2021}.
In the encoding into the wrinkle structure, we first carried out the conversion into the unidirectional stretching field $\bm{s}(u,v)$ by the local/global algorithm.
The result is shown in Fig.~\ref{fig:demonstration}(b).
The line segments represent the direction of $\bm{s}(u,v)$, and the color represents its magnitude.
Next, we computed the wrinkle phase $\phi(u,v)$ by the stripe pattern method, with the wrinkle amplitude set to $A=0.02$. The result is shown in Fig.~\ref{fig:demonstration}(c).
We then obtained the wrinkle height function $w(u,v)$ by Eqs.~(\ref{eq:w_phi}) and~(\ref{eq:f_phi}), yielding the wrinkle structures shown in Fig.~\ref{fig:demonstration}(d).
The color represents the deviation from isometry between the wrinkle structure and the target surface shape, as defined by Eq.~(\ref{eq:discrepancy_metric}).
The color bar is on a logarithmic scale.
The results of applying the amplitude optimization to these wrinkle structures are shown in Fig.~\ref{fig:demonstration}(e).
The amplitude optimization alleviated the deviation from isometry, particularly in the vicinity of the singular points.
Finally, we performed an unfolding simulation of the obtained wrinkle structures and obtained the unfolding-simulation results shown in Fig.~\ref{fig:demonstration}(f).
The unfolding-simulation results superimposed on the target surfaces are shown in Fig.~\ref{fig:demonstration}(g).
The mesh resolutions, iteration counts, pressure-loading schedule, and boundary conditions used in the numerical construction and unfolding simulations are described in \textit{SI Appendix}.

\begin{figure*}[htbp]
    \centering
    \includegraphics[width=\linewidth]{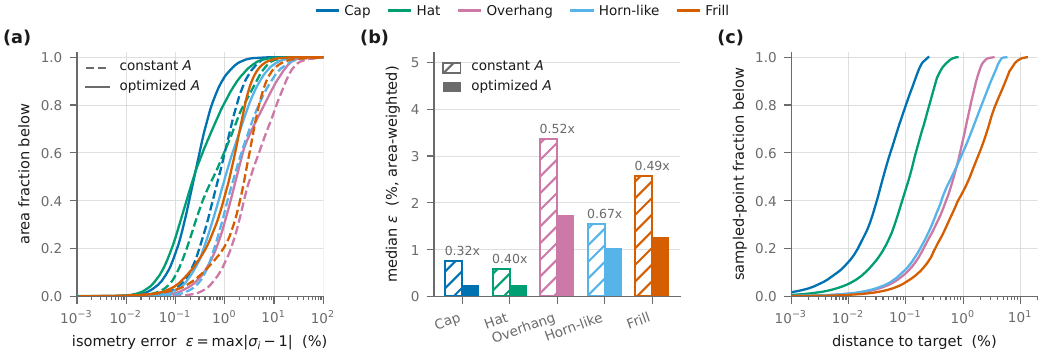}
    \caption{Quantitative evaluation of the designed wrinkle structures. (a) Area-weighted cumulative distribution of the per-triangle isometry error $\varepsilon = \max(|\sigma_1 - 1|, |\sigma_2 - 1|)$ (Eq.~(\ref{eq:eps_metric})) between the wrinkle structure and the target surface, before (dashed) and after (solid) the amplitude optimization. (b) Area-weighted median of $\varepsilon$ for each target shape before and after the amplitude optimization; the number above each pair is the ratio of the two medians. (c) Cumulative distribution of the unsigned distance between the unfolded structure and the target surface after rigid registration, sampled on both surfaces and normalized by the diagonal of the target's bounding box.}
    \label{fig:quantitative}
\end{figure*}

Figure~\ref{fig:quantitative} quantifies these results.
The area-weighted median of the strain-like error $\varepsilon$ (Eq.~(\ref{eq:eps_metric})) is 0.6--3.4\% for the constant-amplitude structures and 0.2--1.7\% after the amplitude optimization; the optimization reduces the median error to 32--67\% of its pre-optimization value across the five target shapes (Fig.~\ref{fig:quantitative}(a) and (b)), with the residual error concentrated in small regions around the phase singularities.
The designed wrinkle structures are thus nearly isometric to their targets, with typical principal-stretch deviations of order 1\% or less.
These error levels are, moreover, not specific to the amplitude chosen for the demonstrations: at the larger amplitude used for 3D printing below ($A=0.12$), the median errors are comparable (\textit{SI Appendix}, Table~S2).
The unfolded shapes agree with the targets to within a mean distance of 0.06--2.1\% of the target bounding-box diagonal (Fig.~\ref{fig:quantitative}(c)); the agreement is closest for the cap and hat (0.06\% and 0.15\%) and loosest for the frill (2.1\%), whose crest placement is not determined by the intrinsic metric (see Discussion).
Note that the unfolding simulations were run on meshes coarsened for computational cost (\textit{SI Appendix}), so these distances include remeshing error as well as unfolding error.
Together, these measurements confirm quantitatively that the inverse origami principle yields wrinkle structures that are nearly isometric to the target surfaces and that can be unfolded mechanically into three-dimensional shapes closely resembling the targets.

\begin{figure*}[htbp]
    \centering
    \includegraphics[width=\linewidth]{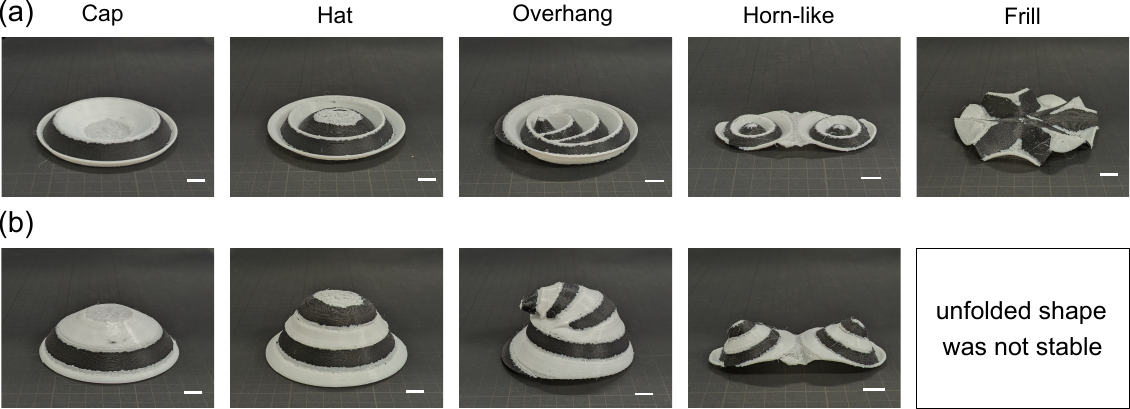}
    \caption{Result of the physical implementation by 3D printing. (a) The 3D-printed wrinkle structures. (b) Unfolded shapes. The white parts were printed with TPU and the black parts with PETG. The unfolded state of the frill structure was not mechanically stable, so no unfolded photograph is shown for it (see Discussion). Scale bars, 10~mm.}
    \label{fig:3dprint}
\end{figure*}

Furthermore, we carried out physical fabrication by 3D printing for all five target shapes. Because 3D printing is difficult when the wrinkle amplitude is too small, we printed wrinkle structures designed with \(A=0.12\); as noted above, the metric error of the designs at this amplitude is comparable to that at \(A=0.02\) (\textit{SI Appendix}, Table~S2). Photographs of the 3D-printed wrinkle structures and of the structures after manual unfolding are shown in Fig.~\ref{fig:3dprint}. By manually unfolding the specimens, we obtained three-dimensional shapes that reproduced the characteristic features of the respective targets, including the overhanging region of the overhang shape; the exception was the frill, whose unfolded state was not mechanically stable and reverted toward the folded configuration upon release (see Discussion). These results confirm that the designed wrinkle structures can be physically fabricated and, with the exception noted, unfolded into shapes close to the target surfaces.

\section*{Discussion}
In this study, inspired by insect exoskeletal morphogenesis, we formulated an inverse-origami problem in which a final three-dimensional surface is encoded into a preformed wrinkle structure. By representing wrinkle unfolding as a unidirectional stretching field, we developed a method for deriving a wrinkle structure from a prescribed target surface. The numerical demonstrations illustrate the applicability of the geometric construction, while the 3D-printed structures connect the resulting designs to physical fabrication.

Applied to the insect, this principle specifies the geometric information that a wrinkle structure encoding a target shape would have to carry, rather than the developmental mechanism by which that information is implemented. The unidirectional stretching field is an effective geometric quantity and should not be identified directly with cell-division orientation, tissue growth, or stress. At the beetle-horn tip, cell-division orientation appears to play only a minor role, whereas spatial variation in division frequency, and hence areal change, is thought to influence the resulting surface shape~\cite{Adachi2018,Morikawa2023}. Our framework therefore provides a reference for investigating how cellular activities and mechanical interactions generate a wrinkle structure with the required geometry.

The horn-like target produced two nearly concentric wrinkle patterns separated by a hyperbola-like boundary, qualitatively resembling the wrinkle structure at the tip of the beetle horn primordium in Fig.~\ref{fig:beetle_horn}(a)~\cite{Matsuda2017,Matsuda2021}. Because the target was a CAD model that merely mimicked the horn tip and the comparison was qualitative, this resemblance should not be regarded as biological validation. Rather, it provides a hypothesis to be tested using measured target shapes and quantitative comparisons with actual wrinkle patterns. Ridge terminations and changes in wrinkle orientation may also be interpreted as singularities or topological defects of the orientation field, a viewpoint developed for wrinkle patterns in the physics of thin sheets~\cite{Aharoni2017}; determining how their positions and indices relate to branching and local surface geometry is an important future problem.

The scope and exactness of the theory require several qualifications. The construction is exact only when the target surface admits a parameterization whose Jacobian has singular values $\sigma_1\geq1$ and $\sigma_2=1$, and when the associated wave-number field $\bm{k}=\frac{\pi}{2A}\sqrt{\sigma_1^2-1}\,\bm{v}_1$ is globally compatible with a phase field, such that $\bm{k}=\nabla\phi$ after a consistent choice of orientation and with the appropriate compatibility conditions around singularities and non-simply-connected regions. Under these conditions and for constant wrinkle amplitude $A$, the constructed wrinkle surface has the same induced metric as the target surface wherever $f$ is differentiable; along the crease curves, the isometry is understood in the piecewise-smooth sense. For general target surfaces, however, the local/global algorithm only minimizes the deviation from the required singular-value condition, and the stripe-pattern method approximates the phase variations prescribed by $\bm{k}$ when no globally compatible phase exists. The amplitude optimization further modifies the constant-amplitude construction throughout the domain, most strongly near the singularities. The final numerical wrinkle structures should therefore be regarded as approximate solutions and assessed using metric-error measures.

The framework is also restricted to wrinkles that can be represented as height functions over a plane. This approximation is appropriate for the gently curved, nearly planar horn-tip region visible in Fig.~\ref{fig:beetle_horn}(a), whereas the complete horn primordium, including the shaft, forms a wrinkle structure on an approximately mushroom-shaped surface~\cite{Matsuda2021}. Because insect wrinkle structures need not lie on nearly planar substrates, extending the inverse origami principle to wrinkles on general curved surfaces is an important future problem. In addition, the wrinkle direction becomes ill-conditioned where the stretch ratio approaches one, as seen near the pole of the hat shape in Fig.~\ref{fig:demonstration}(d). Extending the framework to curved substrates while treating singularities, variable amplitude, and field regularity consistently remains an important challenge.

The frill shape illustrates a further distinction between intrinsic geometry and extrinsic embedding. Even when the metric is rotationally symmetric, the placement and number of crests and troughs are not uniquely determined by that metric, but can depend on bending stiffness, boundary conditions, and discretization. Because our unfolding simulation includes bending energy and uses a coarsened mesh, it should be interpreted primarily as a test of mechanical feasibility and a visualization of unfolding rather than as evidence of isometry; isometry is more appropriately evaluated from the metric error.

Viewed from an engineering perspective, the present method can be regarded as a strategy for fabricating prescribed curved surfaces. Two related classes of approaches are origami design methods, which create target shapes by folding an inextensible flat sheet~\cite{Demaine2017,Dudte2016,Sardas2025,Zhao2023}, and programmable materials, which shape a planar sheet through an encoded local metric or deformation field~\cite{Aharoni2018,Panetta2021,Siefert2020,KonakovicLukovic2018}. Our method shares with the latter the first step of representing the target surface as a unidirectional stretching field on a plane. Its distinctive second step is to convert this field into a preformed wrinkle surface that, under the compatibility conditions described above, is isometric to the target. Unfolding the wrinkles then realizes the effective stretch geometrically, without material extension. In this respect, the method is origami-like; unlike conventional origami, however, the initial configuration is a wrinkled surface that is represented over a planar domain but is intrinsically isometric to the target, rather than an intrinsically flat sheet. Pop-up printing shares the fabrication strategy of producing a structure in a folded configuration and deploying it into its final form by unfolding~\cite{Noma2020}. In that approach, however, the crease pattern is prescribed by the user rather than derived from the target geometry. Our contribution is a geometric theory for deriving the wrinkle geometry, including its crease pattern, from a prescribed target surface.

The 3D-printed structures provide a proof of concept that wrinkle structures obtained by this geometric design procedure can be fabricated and unfolded into shapes reproducing the characteristic features of their targets. They do not, however, establish isometry of the physical specimens or the general stability of the unfolded state. In the zero-thickness theory, inverting a crease costs no stretching; in a sheet of finite thickness, by contrast, inversion induces through-thickness strain and bending energy, and the resulting forces distort the unfolded shape. Following the multimaterial strategy of pop-up printing~\cite{Noma2020}, we therefore printed the regions that need not deform in a stiff material that resists this distortion. The stability observed in the present specimens was neither an outcome of the geometric design nor systematically evaluated. More generally, restoring forces associated with unfolding may return the structure toward its folded state~\cite{Noma2020}; the printed frill specimen exhibited exactly this behavior, its unfolded state reverting toward the folded configuration upon release. Robust self-deploying structures will therefore require the wrinkle geometry to be designed together with material stiffness, thickness, boundary constraints, and loading protocols.

By distinguishing the geometry that must be encoded from the mechanisms that implement it, the inverse origami principle provides a common framework for designing deployable structures and programmable materials, and for generating testable hypotheses about insect morphogenesis.

% --- PNAS: Materials and Methods ---
\matmethods{%
The numerical construction pipeline---the local/global parameterization, the stripe-pattern phase computation, the remeshing along crease lines, and the amplitude optimization---was implemented in custom C++ code; the energy functions and update rules of each stage, together with the mesh resolutions and iteration counts used for the demonstrations, are given in \textit{SI Appendix}.
The unfolding simulations minimize the discrete shell energy described in the main text by simulated annealing, on meshes coarsened to $10^4$ triangles; the energy functional, the annealing schedule, and all parameter values are listed in \textit{SI Appendix}, Table~S1.
The isometry error was computed from the per-triangle singular values of the map from the reference surface to the designed structure and summarized by area-weighted statistics, and the agreement between the unfolded shapes and the targets was measured by point-to-surface distances after rigid ICP registration (Open3D); the exclusion criteria, registration settings, and sampling parameters of both evaluations are given in \textit{SI Appendix}.
The wrinkle structures were fabricated on a Prusa XL multimaterial 3D printer in TPU and PETG with locally thinned crease bands; the geometry processing and fabrication parameters are given in \textit{SI Appendix}.
}
\showmatmethods{}

% --- PNAS: Data availability statement (to be completed) ---
\dataavail{All code and data needed to reproduce the results will be deposited in Zenodo and made publicly available upon publication. These include the code for the design pipeline, the unfolding simulations, and the quantitative evaluation, together with the target surface meshes, the designed wrinkle structures, the simulated unfolded shapes, and the files used for 3D printing. These materials are available to editors and reviewers upon request during peer review.}

\acknow{This work was supported by MEXT KAKENHI (Grant Nos. 20H05947 and 26H01809 to Y.I., and 23H04316 and 25H01364 to N.S.); JSPS KAKENHI (Grant Nos. 24K23206 and 25K21510 to K.M.); and the Engineering Research Grant from the Hattori Hokokai Foundation (to K.M.).}

\showacknow{} % Display the acknowledgments section

% \bibsplit[N] starts the reference list as a full-width-separated block and breaks to the
% right column after the N-th reference. N is chosen so that the two columns are balanced.
\bibsplit[17]

% pnas-new.cls automatically selects pnas-new.bst, so no \bibliographystyle is needed.
\bibliography{reference}

\input{arxiv_si_begin}

\input{si_body}

\end{document}

%% file: arxiv_preamble.tex
\makeatletter

\setboolean{displaywatermark}{false}

\fancypagestyle{firststyle}{%
  \fancyhf{}%
  \fancyfoot[C]{\footerfont\thepage}}
\renewcommand{\@maketitle}{%
  {\raggedright\baselineskip=24pt\titlefont \@title\par}%
  \vskip10pt
  {\raggedright \@author\par}
  \vskip8pt
  {\raggedright \datesfont \@ifundefined{@dates}{}{\@dates}\par}
  \vskip12pt
  \@additionalelement
  \vskip\pnas@vertadjust}

\renewcommand{\dates}[1]{}
\renewcommand{\doi}[1]{}

\newcommand{\arxivbibsplit}{17}
\renewcommand{\bibsplit}[1][]{%
  \begin{strip}\vspace*{-12pt}\end{strip}\setcounter{BalanceAtReference}{\arxivbibsplit}}

\usepackage{placeins}

\makeatother

%% file: arxiv_si_begin.tex
% =====================================================================
% Start of the SI Appendix appended to the arXiv version.
% build.sh inputs this file and then si_body.tex (the body of SI.tex)
% after the reference list.
% =====================================================================
\clearpage
\onecolumn

% S-prefixed equation, figure, and table numbers, as in the PNAS SI.
% \theH... keep the hyperref anchors distinct from those of the main text.
\setcounter{equation}{0}
\setcounter{figure}{0}
\setcounter{table}{0}
\renewcommand{\theequation}{S\arabic{equation}}
\renewcommand{\thefigure}{S\arabic{figure}}
\renewcommand{\thetable}{S\arabic{table}}
\renewcommand{\theHequation}{S\arabic{equation}}
\renewcommand{\theHfigure}{S\arabic{figure}}
\renewcommand{\theHtable}{S\arabic{table}}

{\raggedright\titlefont SI Appendix\par}
\vskip12pt

%% file: si_body.tex
This Supporting Information provides the derivation of the isometry condition, the
implementation details of the numerical construction pipeline, the energy functional used
in the unfolding simulation, the quantitative evaluation procedures, and the 3D-printing
parameters, all of which are summarized in the main text. Equation, figure, and table
numbers prefixed with ``S'' refer to this document; unprefixed equation numbers refer to
the main text.

\section*{Derivation of the isometry condition}
Here we derive the isometry condition of the main text (Eq.~\ref{eq:isometry_condition})
by computing and equating the pullback metrics of the target surface and of the wrinkle
structure.

The target surface $S$ is endowed with the metric $g_{E^3}=\sum_{i=1}^3 (dX^i)^2$ induced
from the Euclidean metric, and its pullback by the parameterization $\varphi$ is in
general written as $\varphi^*g_{E^3}=E\,du^2 + 2F\,du\,dv + G\,dv^2$. Noting that, using
the Jacobian matrix $J_\varphi$ of $\varphi$, it can be written as
\begin{equation}
    \begin{pmatrix} E & F \\ F & G \end{pmatrix} = J_\varphi^T J_\varphi,
\end{equation}
and that the parameterization is such that $\sigma_2 = 1$, the pullback can be written as
\begin{equation}
    \begin{split}
    \varphi^*g_{E^3} &= \begin{pmatrix} du & dv \end{pmatrix} J_\varphi^T J_\varphi \begin{pmatrix} du \\ dv \end{pmatrix} \\
    &= \begin{pmatrix} du & dv \end{pmatrix} V\begin{pmatrix} (\sigma_1)^2 & 0 \\ 0 & 1 \end{pmatrix} V^T \begin{pmatrix} du \\ dv \end{pmatrix}.
    \end{split}
\end{equation}
Here, writing $\bm{v}_1 = \begin{pmatrix} \cos \alpha & \sin \alpha \end{pmatrix}^T$ and $\bm{v}_2 = \begin{pmatrix} \mp\sin \alpha & \pm\cos \alpha \end{pmatrix}^T$, this becomes
\begin{equation}
    \begin{split}
    \varphi^*g_{E^3} &= \begin{pmatrix} du & dv \end{pmatrix} \begin{pmatrix} \cos \alpha & \mp\sin \alpha \\ \sin \alpha & \pm\cos \alpha \end{pmatrix} \begin{pmatrix} (\sigma_1)^2 & 0 \\ 0 & 1 \end{pmatrix} \begin{pmatrix} \cos \alpha & \sin \alpha \\ \mp\sin \alpha & \pm\cos \alpha \end{pmatrix} \begin{pmatrix} du \\ dv \end{pmatrix} \\
    &= \begin{pmatrix} du & dv \end{pmatrix} \begin{pmatrix} 1 + ((\sigma_1)^2 - 1) \cos^2 \alpha & ((\sigma_1)^2 - 1) \cos \alpha \sin \alpha \\ ((\sigma_1)^2 - 1) \cos \alpha \sin \alpha & 1 + ((\sigma_1)^2 - 1) \sin^2 \alpha \end{pmatrix} \begin{pmatrix} du \\ dv \end{pmatrix}.
    \end{split}
\end{equation}

On the other hand, for the wrinkle structure $\varphi_\mathrm{W}(u,v) = (u, v, w(u,v))$,
using the height function of Eqs.~\ref{eq:w_phi} and~\ref{eq:f_phi} of the main text, the
metric induced from the Euclidean metric is
\begin{equation}
    \begin{split}
    \varphi_\mathrm{W}^*g_{E^3} &= du^2 + dv^2 + dw^2 = du^2 + dv^2 + \left[d\left(A f(\phi(u,v))\right)\right]^2 \\
    &= du^2 + dv^2 + A^2\left(\frac{df}{d\phi}\right)^2 \left(\left(\frac{\partial \phi}{\partial u}\right)^2 du^2 + 2\frac{\partial \phi}{\partial u}\frac{\partial \phi}{\partial v} du dv + \left(\frac{\partial \phi}{\partial v}\right)^2 dv^2\right) \\
    &= \begin{pmatrix} du & dv \end{pmatrix} \begin{pmatrix} 1 + \left(\frac{2A}{\pi}\right)^2\left(\frac{\partial \phi}{\partial u}\right)^2 & \left(\frac{2A}{\pi}\right)^2\frac{\partial \phi}{\partial u}\frac{\partial \phi}{\partial v} \\ \left(\frac{2A}{\pi}\right)^2\frac{\partial \phi}{\partial u}\frac{\partial \phi}{\partial v} & 1 + \left(\frac{2A}{\pi}\right)^2\left(\frac{\partial \phi}{\partial v}\right)^2 \end{pmatrix} \begin{pmatrix} du \\ dv \end{pmatrix}.
    \end{split}
\end{equation}
Since the condition for the target surface and the wrinkle structure to be isometric is
that these two metrics coincide, we obtain
\begin{align}
    \label{si:eq:phi_u}
    \frac{\partial\phi}{\partial u} &= \pm\frac{\pi}{2A}\sqrt{(\sigma_1)^2 - 1} \cos \alpha \\
    \label{si:eq:phi_v}
    \frac{\partial\phi}{\partial v} &= \pm\frac{\pi}{2A}\sqrt{(\sigma_1)^2 - 1} \sin \alpha \\
    \label{si:eq:phi_u_v_product}
    \frac{\partial\phi}{\partial u}\frac{\partial\phi}{\partial v} &= \left(\frac{\pi}{2A}\right)^2((\sigma_1)^2 - 1) \cos \alpha \sin \alpha
\end{align}
Equation~(\ref{si:eq:phi_u_v_product}) holds provided that the double signs in
Eqs.~(\ref{si:eq:phi_u}) and~(\ref{si:eq:phi_v}) are taken in the same order. Therefore,
the isometry condition is
\begin{equation}
    \nabla\phi = \pm\frac{\pi}{2A}\sqrt{(\sigma_1)^2 - 1} \begin{pmatrix} \cos \alpha \\ \sin \alpha \end{pmatrix} = \pm\frac{\pi}{2A}\sqrt{(\sigma_1)^2 - 1}\;\bm{v}_1,
\end{equation}
which is Eq.~\ref{eq:isometry_condition} of the main text.

\section*{Numerical construction of the wrinkle structure}

\subsection*{Parameterization for the unidirectional stretching field}
The parameterization $\varphi: D \to S$ whose per-triangle Jacobian is constrained to the
set $M$ of allowed linear transformations (Eq.~\ref{eq:allowed_linear_transformations} of
the main text) is computed with the local/global algorithm~\cite{Liu2008}, an extension of
as-rigid-as-possible surface modeling~\cite{Sorkine2007}. We find the vertex positions
$\{\bm{u}_i\}=\{\bm{u}_1,\dots,\bm{u}_{N_\mathrm{v}}\}$ and the per-triangle linear
transformations $\{L_i\}=\{L_1,\dots,L_{N_\mathrm{f}}\}$ that minimize
\begin{equation}
    \label{si:eq:local_global_energy}
    E(\{\bm{u}_i\}, \{L_i\}) = \sum_{i=1}^{N_\mathrm{f}} A_i\|J_i(u,v) - L_i\|^2_F,
\end{equation}
where $A_i$ is the area of the $i$-th triangle and $\|\cdot\|_F$ is the Frobenius norm.
The algorithm minimizes Eq.~(\ref{si:eq:local_global_energy}) by alternating a local phase,
which fixes $\{\bm{u}_i\}$ and finds $\{L_i\}$, and a global phase, which fixes $\{L_i\}$
and finds $\{\bm{u}_i\}$; the initial values are obtained by conformal
parameterization~\cite{Eck1995, Levy2002}. In the local phase, from the singular value
decomposition of the Jacobian matrix at the current $\{\bm{u}_i\}$,
\begin{equation}
    J_i(u,v) = U_i\begin{pmatrix} \sigma_{1,i} & 0 \\ 0 & \sigma_{2,i} \\ 0 & 0 \end{pmatrix}V_i^T,
\end{equation}
we update $\{L_i\}$ as
\begin{equation}
    L_i = \begin{cases} U_i\begin{pmatrix} \sigma_{1,i} & 0 \\ 0 & 1 \\ 0 & 0 \end{pmatrix}V_i^T & \text{if } \sigma_{1,i} \geq 1, \\[6pt] U_i\begin{pmatrix} 1 & 0 \\ 0 & 1 \\ 0 & 0 \end{pmatrix}V_i^T & \text{if } \sigma_{1,i} < 1. \end{cases}
\end{equation}
In the global phase, for the energy at the current $\{L_i\}$ we solve
\begin{equation}
    \frac{\partial E(\{\bm{u}_i\}, \{L_i\})}{\partial \bm{u}_j} = 0 \quad (j = 1, 2, \dots, N_\mathrm{v})
\end{equation}
to update $\{\bm{u}_i\}$. Repeating these phases yields the parameterization $\{\bm{u}_i\}$,
and from $\sigma_{1,i}$ and $V_i=\begin{pmatrix} \bm{v}_{1,i} & \bm{v}_{2,i} \end{pmatrix}$
in the resulting $\{L_i\}$, the unidirectional stretching field
$\bm{s}_i=\sigma_{1,i}\bm{v}_{1,i}$ is obtained for each triangle (Eq.~\ref{eq:s_param} of
the main text).

\subsection*{Constructing the phase field by the stripe pattern method}
From the unidirectional stretching field $\bm{s}_i$, we compute the wave-number vector
field $\bm{k}_i$ for each triangle (Eq.~\ref{eq:k_s} of the main text) and seek a phase
$\phi$ with $\nabla\phi=\bm{k}$ (Eq.~\ref{eq:k_phi} of the main text). Because $\bm{k}_i$ is
in general not curl-free, no globally continuous scalar function has gradient $\bm{k}_i$; we
therefore use the stripe pattern method~\cite{Knoppel2015}, which represents the phase as
the argument $\phi=\arg\psi$ of a complex function $\psi$. Numerically, for each edge
$\bm{e}_{ij}=\bm{u}_j-\bm{u}_i$ of the triangle mesh, using the change in argument along the
edge $\omega_{ij}=\bm{k}_{ij}\cdot\bm{e}_{ij}$, where $\bm{k}_{ij}$ is the average of the
wave-number vectors at vertices $i$ and $j$, we determine
$\{\psi_i\}=\{\psi_1,\dots,\psi_{N_\mathrm{v}}\}$ minimizing
\begin{equation}
    E(\{\psi_i\}) = \sum_{ij\in\mathrm{edges}} w_{ij} \left| \psi_j - e^{\iota\omega_{ij}} \psi_i \right|^2,
\end{equation}
where $\iota$ is the imaginary unit, $w_{ij}=(\cot \beta_{ij} + \cot \beta_{ji})/2$, and
$\beta_{ij}, \beta_{ji}$ are the angles opposite edge $\bm{e}_{ij}$. Writing
$\bm{\psi}=(\psi_1,\dots,\psi_{N_\mathrm{v}})^T\in\mathbb{C}^{N_\mathrm{v}}$, this energy is
the Hermitian quadratic form $E(\{\psi_i\})=\bm{\psi}^\ast A\bm{\psi}$, where $A$ is a
Hermitian positive-definite matrix containing $e^{\iota\omega_{ij}}$ (see~\cite{Knoppel2015}
for its construction). To avoid the trivial solution $\bm{\psi}=\bm{0}$, we impose the $L^2$
normalization $\bm{\psi}^\ast B\bm{\psi}=1$ via the discrete mass matrix $B$; minimizing $E$
under this constraint is equivalent to the generalized eigenvalue problem
\begin{equation}
    A\bm{\psi}=\lambda B\bm{\psi},
\end{equation}
whose eigenvector for the smallest eigenvalue, computed by the inverse power method, gives
$\phi_i=\arg\psi_i$. The height $w_i$ at each vertex then follows from Eqs.~\ref{eq:f_phi}
and~\ref{eq:w_phi} of the main text. Because the phase isolines $\phi=0,\pi$ (the crease
lines) may cross triangle interiors, we remesh by adding edges so that the crease lines lie
on the edges of the triangle mesh, completing the mesh of the wrinkle structure.

\subsection*{Amplitude optimization at singular points}
To resolve the height discontinuities that appear at singular points of the wave-number
field, we adjust the wrinkle amplitude with the same as-rigid-as-possible
framework~\cite{Sorkine2007} used for the parameterization. Whereas the parameterization
sought the $(u,v)$ coordinates $\{\bm{u}_i\}$, the amplitude adjustment seeks the height
coordinates $\{w_i\}=\{w_1,\dots,w_{N_\mathrm{v}}\}$; and whereas the parameterization
constrained the per-triangle deformation to $M$
(Eq.~\ref{eq:allowed_linear_transformations} of the main text), the amplitude adjustment
restricts it to rigid rotations, so as to remain isometric to the target surface:
\begin{equation}
    M^\mathrm{rigid} = \left\{ R \in \mathbb{R}^{3 \times 3} \mid R = U\begin{pmatrix} 1 & 0 & 0 \\ 0 & 1 & 0 \\ 0 & 0 & 0 \end{pmatrix}V^T \right\},
\end{equation}
where $U$ and $V$ are orthogonal matrices. ($R$ is $3\times3$ because both the reference
shape and the current shape lie in three-dimensional space, and two entries equal $1$
because the per-triangle deformation has two components.) We minimize
\begin{equation}
    E(\{w_i\}, \{R_i\}) = \sum_{i=1}^{N_\mathrm{f}} A_i\|J_i(w) - R_i\|^2_F,
\end{equation}
alternating a local phase, in which for the singular value decomposition
$J_i(w)=U_i\Sigma_i V_i^T$ we set
\begin{equation}
    R_i = U_i\begin{pmatrix} 1 & 0 & 0 \\ 0 & 1 & 0 \\ 0 & 0 & 0 \end{pmatrix}V_i^T,
\end{equation}
and a global phase, in which we solve
\begin{equation}
    \frac{\partial E(\{w_i\}, \{R_i\})}{\partial w_j} = 0 \quad (j = 1, 2, \dots, N_\mathrm{v})
\end{equation}
to update $\{w_i\}$. This iteration yields a wrinkle structure with position-dependent
amplitude $A(u,v)=w(u,v)/f(\phi(u,v))$, which alleviates the discontinuity of the height
function at the singular points.

\section*{Energy functional for the unfolding simulation}
Mimicking the pressure-driven unfolding of the beetle horn, the unfolding simulation is a
mechanical computation in which internal pressure is applied while isometry is maintained. It is
performed as a minimization, by simulated annealing, of the energy defined below
(see~\cite{Morikawa2023} for details):
\begin{equation}
    E=E^{\mathrm{Area}} + E^{\mathrm{IA}} + E^{\mathrm{DA}} + E^{\mathrm{Vol}},
\end{equation}
\begin{equation}
    E^{\text{Area}}=\sum_i^{N_{\mathrm{f}}} \frac{1}{2} k_{\text{Area}} \frac{\left(A_i-A_i^{\mathrm{init}}\right)^2}{A_i^{\mathrm{init}} \cdot A_{\text{total}}^{\mathrm{init}}}, \quad A_{\text{total}}^{\mathrm{init}}=\sum_i^{N_{\mathrm{f}}} A_i^{\mathrm{init}},
\end{equation}
\begin{equation}
    E^{\mathrm{IA}}=\sum_i^{N_{\mathrm{f}}} \frac{1}{2} k_{\mathrm{IA}} \sum_j^3\left(\frac{\theta_{i, j}^{\mathrm{IA}}}{\theta_{i, j}^{\mathrm{IA}, \mathrm{init}}}-1\right)^2 \cdot \frac{\frac{1}{2}\left(L_{i, j, 1}^{\mathrm{init}}+L_{i, j, 2}^{\mathrm{init}}\right)}{P_{\mathrm{total}}^{\mathrm{init}}}, \quad P_{\mathrm{total}}^{\mathrm{init}}=\sum_i^{N_{\mathrm{f}}} P_i^{\mathrm{init}},
\end{equation}
\begin{equation}
    E^{\mathrm{DA}}=\sum_i^{N_\mathrm{e}} k_{\mathrm{DA}} \frac{3\left(L_i^{\mathrm{init}}\right)^2}{A_{i_1}^{\mathrm{init}}+A_{i_2}^{\mathrm{init}}}\left(2 \tan \frac{\theta_i^{\mathrm{DA}}}{2}\right)^2,
\end{equation}
\begin{equation}
    E^{\mathrm{Vol}}= \frac{1}{2} k_{\mathrm{Vol}} \left(\frac{V}{V^\mathrm{tgt}}-1\right)^2.
\end{equation}
Here, $E^{\mathrm{Area}}$ and $E^{\mathrm{IA}}$ represent the deviations of the area $A_i$ of
each triangle $i$ and of the interior angles $\theta^{\mathrm{IA}}_{i,j}\ (j=1,2,3)$,
respectively, from their initial-shape values $A_i^{\mathrm{init}}$ and
$\theta^{\mathrm{IA},\mathrm{init}}_{i,j}$, and act so as to maintain
isometry~\cite{Grinspun2003}. $E^{\mathrm{DA}}$ represents the deviation from flatness of the
dihedral angle $\theta_i^\mathrm{DA}$ between the two faces $i_1, i_2$ sharing each edge $i$,
and corresponds to a bending energy~\cite{Tamstorf2013}; this term drives the unfolding of
the wrinkles. In the definition of $E^{\mathrm{IA}}$, $L_{i,j,k}^{\mathrm{init}}\ (k=1,2)$ is
the initial length of the edge $k$ sharing vertex $j$ of triangle $i$, and
$P_i^{\mathrm{init}}$ is the initial perimeter of triangle $i$; in the definition of
$E^{\mathrm{DA}}$, $L_i^{\mathrm{init}}$ is the initial length of edge $i$ and $N_\mathrm{e}$
is the number of interior edges of the mesh. $E^{\mathrm{Vol}}$ is the internal-pressure
energy that drives the unfolding, bringing the lumen volume $V$ close to the target volume
$V^\mathrm{tgt}$~\cite{Morikawa2023}. To exclude the stretching effect of the internal
pressure, $E^{\mathrm{Vol}}$ acts only during the first $n_1$ steps, with $k_{\mathrm{Vol}}=0$
thereafter. The simulated annealing was performed with $n_1$ steps at high temperature
(inverse temperature $\beta=\beta_1$) and $n_2$ steps at low temperature
($\beta=\beta_2>\beta_1$)~\cite{Morikawa2023}. The parameters are listed in
Table~\ref{si:tab:parameters}.

\begin{table}[htbp]
    \centering
    \caption{Parameters of the unfolding simulation.}
    \label{si:tab:parameters}
    \begin{tabular}{lll}
        \toprule
        Symbol& Value& Description\\
        \midrule
        $\beta_1$ & $1.0\times10^3$ & Inverse temperature for the first $n_1$ steps \\
        $\beta_2$ & $1.0\times10^6$ & Inverse temperature for the second $n_2$ steps \\
        $k_\mathrm{Area}$ & $1.0\times10^4$ & Constant of area energy \\
        $k_\mathrm{IA}$ & $1.0\times10^5$ & Constant of interior angle energy \\
        $k_\mathrm{DA}$ & $2.0\times10^{-2}$ & Constant of dihedral angle energy \\
        $k_\mathrm{Vol}$ & $2.0\times10^{3}$ & Constant of volume energy for the first $n_1$ steps \\
        $V^\mathrm{tgt}$ & $3.0$ & Target value of volume \\
        \bottomrule
    \end{tabular}
\end{table}

\section*{Numerical parameters for the demonstrations}
The cap, hat, overhang, horn-like, and frill target meshes in Fig.~\ref{fig:demonstration} of the main text consisted of $N_\mathrm{f}=30415$, $30258$, $39756$, $30328$, and $34077$ triangles, respectively. The local/global algorithm was run for $2,000$ steps, and the amplitude optimization was run for $200$ steps. To reduce the computational cost of the unfolding simulations, each mesh was reduced to $10,000$ triangles using Isotropic Explicit Remeshing in MeshLab~\cite{Hoppe1993}. The simulations consisted of $n_1=1.5\times10^5$ steps followed by $n_2=1.35\times10^6$ steps, except for the overhang shape, for which the first phase was run for $n_1=6.0\times10^5$ steps and the second for $n_2=4.0\times10^5$ steps. During the first $n_1$ steps, the boundary was fixed and internal pressure was applied; during the subsequent $n_2$ steps, the pressure was switched off ($k_\mathrm{Vol}=0$) and the computation was performed under free boundary conditions. For the frill shape alone, which is not suited to a fixed boundary and internal pressure, the first $n_1$ steps were also performed with a free boundary and zero internal pressure.

\section*{Quantitative evaluation of the designed structures}

\subsection*{Isometry error}
The per-triangle deviation from isometry (Eqs.~\ref{eq:discrepancy_metric} and~\ref{eq:eps_metric} of the main text) is computed from the Jacobian of the map from each reference (target) triangle to the corresponding triangle of the wrinkle structure, using the one-to-one face correspondence maintained by the construction pipeline. For the constant-amplitude structures, the reference is the subdivided target surface used by the stripe-pattern stage; for the amplitude-optimized structures, it is the reference mesh written by the amplitude-optimization stage. Faces whose reference area is below $10^{-9}$ (a handful of triangles per mesh whose area rounds to zero when the meshes are written to disk) are excluded. All statistics are weighted by reference-triangle area. Because a few tiny, strongly distorted triangles at the phase singularities dominate the area-weighted mean while holding negligible area, we report the area-weighted median and 95th percentile, which are robust to these outliers.

Table~\ref{si:tab:robustness} lists the resulting error statistics at the amplitude used in the numerical demonstrations ($A=0.02$) and at the amplitude used for 3D printing ($A=0.12$). The error levels are comparable at the two amplitudes, both before and after the amplitude optimization, so the reported errors are not specific to the choice of $A$.

\begin{table}[htbp]
    \centering
    \caption{Area-weighted isometry error $\varepsilon$ (\%) of the designed wrinkle structures at the two amplitudes, before (constant $A$) and after amplitude optimization.}
    \label{si:tab:robustness}
    \begin{tabular}{llcccc}
        \toprule
        & & \multicolumn{2}{c}{Median} & \multicolumn{2}{c}{95th percentile} \\
        Shape & $A$ & constant & optimized & constant & optimized \\
        \midrule
        Cap       & 0.02 & 0.76 & 0.24 & 3.65  & 1.34 \\
                  & 0.12 & 0.77 & 0.14 & 3.76  & 0.70 \\
        Hat       & 0.02 & 0.59 & 0.24 & 6.12  & 2.87 \\
                  & 0.12 & 0.81 & 0.33 & 6.75  & 1.56 \\
        Overhang  & 0.02 & 3.36 & 1.74 & 23.39 & 16.42 \\
                  & 0.12 & 4.64 & 2.64 & 32.22 & 15.70 \\
        Horn-like & 0.02 & 1.54 & 1.03 & 13.34 & 7.76 \\
                  & 0.12 & 2.48 & 1.33 & 13.40 & 7.27 \\
        Frill     & 0.02 & 2.57 & 1.27 & 9.76  & 4.76 \\
                  & 0.12 & 4.37 & 1.66 & 12.26 & 5.68 \\
        \bottomrule
    \end{tabular}
\end{table}

\subsection*{Agreement between the unfolded shape and the target}
The final state of each unfolding simulation is rigidly registered onto the target surface by point-to-point iterative closest point (ICP) alignment (Open3D), initialized by aligning the centroids of the two meshes; the free-boundary phase of the simulation permits a global translational drift, which the centroid initialization removes. The down-sampling voxel size and the correspondence-distance threshold are set relative to the target size, at $d/200$ and $d/10$ respectively, where $d$ is the diagonal of the target's bounding box. No scaling is applied, since the parameterization preserves the metric. After registration, $2\times10^5$ points are sampled uniformly on each surface, and the unsigned point-to-surface distances, pooled over both directions, are normalized by $d$. Because the unfolding simulations are run on meshes coarsened to $10^4$ triangles (see above), the reported distances include remeshing error as well as unfolding error.

\section*{Physical implementation by 3D printing}
All five target shapes (Fig.~\ref{fig:demonstration}(a) of the main text) were physically implemented, using the wrinkle structures designed with $A=0.12$. The structures were fabricated using a Prusa XL multimaterial 3D printer (Prusa Research a.s., Czech Republic). Each design was scaled so that the longer side of its bounding rectangle in the $(u,v)$ plane measured 100\,mm. A thickness of 0.4\,mm was assigned to narrow bands along the creases, where the phase satisfies $-\pi < \phi \leq -\pi+\Delta\phi^\mathrm{th},\ -\Delta\phi^\mathrm{th} < \phi \leq \Delta\phi^\mathrm{th},\ \pi-\Delta\phi^\mathrm{th} < \phi \leq \pi\ (\Delta\phi^\mathrm{th}=0.3)$, so that the creases act as flexible hinges; a uniform thickness of 1.4\,mm was assigned everywhere else. PolyFlex TPU95-HE White (Polymaker, China) was used for the regions that invert the surface during unfolding and for the crease bands, whereas Prusament PETG (Prusa Research a.s.) was used for the regions that do not invert. Prusament PLA (Prusa Research a.s.) was used as the support material during printing.

%% Add this line AFTER all your figures and tables.
\FloatBarrier